%% file: pap2.tex
\documentclass[pra,aps,superscriptaddress,secnumarabic,eqsecnum,showpacs]{revtex4}
\usepackage{epsfig,amsmath}
\begin{document}

\title{Quantal Two-Centre Coulomb Problem treated by means of the 
Phase-Integral Method II. Quantization Conditions in the Symmetric Case 
Expressed in Terms of Complete Elliptic Integrals. Numerical Illustration.}

\author{N. Athavan} 
\thanks{Present address: Department of Physics, 
Government Arts College, Ariyalur - 621 713, India.}
\affiliation{Centre for Nonlinear Dynamics, Department of Physics, 
Bharathidasan University, Tiruchirapalli 620 024,India}

\author{N. Fr\"oman} 

\affiliation{Department of Theoretical Physics, University of Uppsala,
Box 803, S-751 05 Uppsala, Sweden}

\author{M. Lakshmanan} 

\affiliation{Centre for Nonlinear Dynamics, Department of Physics, 
Bharathidasan University, Tiruchirapalli 620 024,India}

\begin{abstract}

The contour integrals, occurring in the arbitrary-order phase-integral
quantization conditions given in a previous paper, are in the first-
and third-order approximations expressed in terms of complete elliptic
integrals in the case that the charges of the Coulomb centres are
equal. The evaluation of the integrals is facilitated by the knowledge
of quasiclassical dynamics. The resulting quantization conditions
involving complete elliptic integrals are solved numerically to obtain
the energy eigenvalues and the separation constants of the $1s\sigma$
and $2p\sigma$ states of the hydrogen molecule ion for various values
of the internuclear distance.  The accuracy of the formulas obtained is
illustrated by comparison with available numerically exact results.

\end{abstract}

\maketitle

\section{Introduction}
\label{sec1}
In a previous paper \cite{fro1} the general
two-centre Coulomb problem was treated according to the phase-integral
method, briefly described in the appendix of that paper, and arbitrary-order
quantization conditions were given, valid uniformly for all energies.

The symmetric case, where the charge numbers $Z_1$ and $Z_2$ of the two
Coulomb centres are equal, represents for $Z_1=Z_2=1$ the hydrogen molecule
ion $H_2^+$. This case has for natural reasons been studied more
extensively than the general two-centre problem.
Thus, the $H_2^+$ ion was the subject of treatments already with the aid
of the old quantum theory \cite{hunt} and in the early days of quantum
mechanics \cite{hunt2,bhat}.
The simplicity of the
hydrogen molecule ion grants it an analogous unique position in molecular physics
as the hydrogen atom possesses in atomic physics. For instance, $H_2^+$ plays a role
of fundamental significance in the theory of chemical bonding, and it is also
of importance in the study of stellar atmospheres.

For the background of the present paper we refer to \cite{fro1}. In sec. \ref{sec2} of the present paper
we present the quasiclassical dynamics of the associated particle system
by setting up the equation of motion for a particle moving in the relevant potential
and express the solution in terms of Jacobian elliptic functions.
In sections \ref{sec3} and \ref{sec4}                                    
we express the quantization conditions given in \cite{fro1} in terms
of complete elliptic integrals by using the solution in sec. \ref{sec2}. 
We introduce, analogously 
as Lakshmanan and Kaliappan \cite{lak1}, Lakshmanan {\it et al}. \cite{lak2} 
and Lakshmanan 
{\it et al}. \cite{lak3}, convenient transformations
to elliptic functions, in order to be able to express both the real and the complex
contour integrals, occurring in the quantization conditions, in terms of complete
elliptic integrals. In choosing our transformations we exploit the symmetry of the
functions $R(\eta)$ and $Q^2(\eta)$, introduced in \cite{fro1}, already from the begining of the calculations, since this
is much simpler than to particularize formulas for the general case of arbitrary
nuclear charge numbers $Z_1$ and $Z_2$ to the case $Z_1=Z_2$. Thus we
obtain simple expressions for the quantities 
appearing in the quantization conditions in \cite{fro1}.
It should be remarked that if one particularizes the $\eta$-quantization 
conditions involving complete elliptic integrals that are valid for
arbitrary charge numbers $Z_1$ and $Z_2$ (to be derived in the
subsequent paper \cite{ath3}) to the case $Z_1=Z_2$, one must in general
make further cumbersome transformations in order to bring the
quantization conditions in question into the same form as those obtained
by assuming from the begining
that $Z_1=Z_2$.
The functions $\tilde R(\xi)$ and $\tilde Q^2(\xi)$, introduced in
\cite{fro1},
do not display an analogous symmetry as $R(\eta)$ and $Q^2(\eta)$ when
$Z_1=Z_2$.
The treatment of the $\xi$-equation is thus the same whether $Z_1=Z_2$
or $Z_1\ne Z_2$, and it is analogous to the treatment of the
$\eta$-equation in the general case when $Z_1$ may be different from
$Z_2$, which is treated in the third paper \cite{ath3} in our series of
papers concerning the phase-integral treatment of the quantal
two-centre Coulomb problem.  All the contour integrals needed in the
calculations are expressed in terms of complete elliptic integrals,
corresponding to the use of the first- and third-order phase-integral
approximations. Since complete elliptic integrals can easily be
evaluated by means of standard computer programs, computational
difficulties that may arise in direct numerical calculations
(especially of higher-order terms) are thereby eliminated. One achieves
also the possibility of being able to utilize well-known properties of
complete elliptic integrals, for instance series expansions, for
analytic studies on the basis of the quantization conditions. Some of
the first-order contour integrals have been given in terms of complete
elliptic integrals previously by Strand and Reinhardt \cite{strand},
but to the best knowledge of the present authors, even in the
first-order approximation only partial results have  been given.

The phase-integral formulas derived in this paper are quite flexible
due to the presence of two arbitrary parameters $C$ and $\tilde C$
introduced in the base functions $Q(\eta)$ and $\tilde Q(\xi)$ in eqs.
(3.2b) and (3.2a), respectively, in \cite{fro1}.  In accordance with
the discussion in sec.~3.1 of \cite{fro1} it is appropriate to choose
$C$ and $\tilde C$ such that the first-order results are exactly equal
to the third-order results, and we confirm in sec.~\ref{sec5} the
convenience of this approach by investigating for the $1s\sigma$ and
$2p\sigma$ states of the hydrogen molecule ion the accuracy of our
phase-integral quantization conditions.

\section{quasiclassical dynamics}
\label{sec2}
We have seen in sec. \ref{sec3} of \cite{fro1} that the quantization
conditions require the evaluation of various contour integrals. In the
present section we shall restrict ourselves to the $\eta$-equation. The
shape of the contours depends on the choice of $\Lambda$ and on the
real or complex nature of the zeros of $Q^2(\eta)$; see Figs. 2 - 4 in
\cite{fro1}.

Letting, when $-Q^2(\eta)$ is a double-well potential,
the zeros of $Q^2(\eta)$ be $\eta_1,\eta_2,\eta_3,\eta_4$
when $\Lambda=|m|\ne 0$ and $\eta_2,\eta_3$ when $\Lambda=0$,
we shall evaluate the integrals 
$\alpha=\beta$ and $\bar K$
in the following cases separately.
\begin{enumerate}
\item
$\Lambda=|m|\ne 0$ : 
\begin{enumerate}
\item
Subbarrier case ($\eta_1=-\eta_4 $ and $ \eta_2=-\eta_3$ are real)
\item
Superbarrier case ($\eta_1,\eta_4$ are real and $\eta_2,\eta_3$ are complex conjugate)
\end{enumerate}
\vskip 10pt
\item
$\Lambda=0$: 
\begin{enumerate}
\item
Subbarrier case (only two zeros, $\eta_2 =- \eta_3$; they are real) 
\item
Superbarrier case (only two zeros, $\eta_2$ and $\eta_3$; they are complex conjugate)
\end{enumerate}
\end{enumerate}

In each one of these cases the evaluation of the contour integrals in terms
of complete elliptic integrals is facilitated by the knowledge of the associated
quasiclassical dynamics. In particular the relevant substitution for the integration
variable in the various contour integrals can be obtained with the help
of, for example, Byrd and Friedman \cite{byrd}. However, the meaning of these
substitutions can be attributed to the associated solution of the equation
of motion of the corresponding classical problem as in the case of three-dimensional 
anharmonic oscillators \cite{lak3}.

To illustrate what has been said above, we shall  
consider the subbarrier case of $\Lambda=|m|\ne 0$.
(For the other cases the analysis can be carried out in a similar way.)
The base function $Q(\eta)$ for this case is

\begin{equation}
\label{qqq}
Q(\eta)=p{{[(a^2-\eta^2)(\eta^2-b^2)]^{\frac{1}{2}}}\over {1-\eta^2}},
\end{equation}
where $a=\eta_4=-\eta_1$, $b=\eta_3=-\eta_2$ and $p^2=-\frac{1}{2}Er_{12}^2$.
Let us define a new integration variable $\bar u$ by writing 
\begin{equation}
\bar u=\int^{\eta} {{d\eta}\over {(1-\eta^2)Q(\eta)}};
\end{equation}
then
\begin{equation}
\label{diff}
{{d\eta}\over {d\bar u}}=(1-\eta^2)Q(\eta)=\hat Q(\eta). 
\end{equation}
Before solving (2.3), we shall relate this differential equation to the equation 
of motion for the quasiclassical motion of a particle with a mass $\mu$ in a certain potential 
by differentiating (2.3) with respect to $\bar u$, getting
\begin{equation}
{{d^2\eta}\over {d\bar u^2}}=\frac {1}{2}{{d\hat Q^2}\over {d\eta}}.
\end{equation}
Defining now the ``time'' variable u as
\begin{equation}
u=\bar u\sqrt\mu,
\end{equation}
we obtain the equation of quasiclassical motion
\begin{subequations}
\begin{eqnarray}
\mu {{d^2\eta}\over {du^2}}&=&-{{dU(\eta)}\over {d\eta}},
\end{eqnarray}
\begin{eqnarray}
U(\eta)&=&-\frac{1}{2}\hat Q^2(\eta).
\end{eqnarray}
\end{subequations}
We can consider (2.6a) as representing  the motion of a particle in the
potential $U(\eta)$. We shall now solve the differential equation (\ref{diff}) 
in terms of Jacobian elliptic functions when $Q^2(\eta)$
is given by (\ref{qqq}). The potential $U(\eta)$ is then given by

\begin{eqnarray}
U(\eta) & = &-\frac{1}{2}\hat Q^2(\eta)=-\frac{1}{2}(1-\eta^2)^2Q^2(\eta)\nonumber \\
& = & -\frac{1}{2}p^2(a^2-\eta^2)(\eta^2-b^2)\nonumber\\
& = & \frac{1}{2}p^2(\eta^2-a^2)(\eta^2-b^2),
\end{eqnarray}
and according to (2.3) and (2.1)
\begin{equation}
{{d\eta}\over {d\bar u}}=p\sqrt{(a^2-\eta^2)(\eta^2-b^2)}.
\end{equation}
By solving (2.8) with respect to $\eta$ we obtain
\begin{equation}
\eta(\bar u)={{b}\over{dn\left[\frac{p}{a}(\bar u-\bar u_0)\right]}},
\end{equation}
where $\bar u_0$ is an integration constant, and the modulus $k$ of the Jacobian elliptic
function is given by
\begin{equation}
k^2=1-\frac{b^2}{a^2}. 
\end{equation}
Similarly we can for all other cases  interpret the corresponding
quasiclassical dynamics by the above type of treatment.
\section{Case $\Lambda=|\lowercase{m}|\ne 0$}
\label{sec3}
In order to express the contour integrals, occurring in the quantization
conditions pertaining to the $\xi$-equation and the $\eta$-equation, in terms
of complete elliptic integrals, we transform the integrals in question
from the $\xi$- or $\eta$-plane to another complex plane, the $u$-plane, on
which the Jacobian elliptic functions cn$u$, sn$u$ and dn$u$ are defined.
We make frequent use of formulas in \cite{byrd}.
As mentioned above the transformation in question can be attributed to quasiclassical 
dynamics.

\subsection{Four real zeros of $Q^2(\eta)$ and $\tilde Q^2(\xi)$}
\label{sec:3:1}
\subsubsection{The quantities $\alpha = \beta$ and $\bar K$ pertaining
to the $\eta$-equation: Subbarrier case [Fig. 3(a) in Ref. 1]}
Putting $\eta_4 =-\eta_1 =a$ and $\eta_3=-\eta_2=b$, we have the base function
(2.1), that is,
\begin{equation}
Q(\eta)=p{{[(a^2-\eta^2)(\eta^2-b^2)]^{\frac{1}{2}}}\over {1-\eta^2}}.
\end{equation}
Using the appropriate transformation on p. 54 in \cite{byrd}, 
or equivalently the quasiclassical solution given in sec. II, we put
\begin{equation}
\eta={{b}\over {dn u}} = {{b}\over {(1-k^2sn^2u)^{\frac{1}{2}}}},\hskip 10pt 
k^2=1-{\frac{b^2}{a^2}}.
\end{equation}
Noting that the loop $\eta_2 \rightarrow \eta_1\rightarrow \eta_2$, that is 
$-b\rightarrow -a \rightarrow -b$, in the $\eta$-plane,
denoted by $\Gamma_{-b,-a}$, represents the contour $\Lambda_{\alpha}$
in Fig. 3(a) of \cite{fro1} which corresponds in the $u$ plane to $0\rightarrow 
K\rightarrow 2K$, where
K is the complete elliptic integral of the first kind, we obtain for the 
first-order contribution to $\alpha$ the following transformation
of the original integral in the $\eta$-plane into an integral in the $u$-plane:
\begin{eqnarray}
\alpha^{(1)} & = &{{1}\over{2}}\int_{\Lambda_{\alpha}}Q(\eta)d\eta
= {{1}\over {2}}\int_{\Gamma_{-b,-a}} Q(\eta)d\eta\nonumber \\
& = & {{p}\over{g}}{{1}\over{2}}\int_0^{2K}\left ({{d\eta}\over {du}}\right )^2
{{du}\over {1-\eta^2}} \nonumber\\
& = & p{{k^4b^2}\over {1-b^2}}{{1}\over {2}}\int_0^{2K} {{sn^2u(1-sn^2u)}\over
{(1-k^2sn^2u)(1-\nu^2sn^2u)}}du,
\end{eqnarray}
which, after decomposition of the integrand into partial fractions and use
of recurrence formulas in \cite{byrd}, yields
\begin{subequations}
\label{alp}
\begin{eqnarray}
\alpha^{(1)} & = & {{p}\over {g}}\left[E(k)-\left(1-{{k^2}\over{\nu^2}}\right ) K(k)
+k^2\left (1-{{1}\over {\nu^2}}\right )\Pi(\nu^2,k)\right ],
\end{eqnarray}
\end{subequations}
where
\begin{equation}
\nu^2={{a^2-b^2}\over {a^2(1-b^2)}}={{k^2}\over {1-b^2}}, g={{1}\over {a}}
,k^2={{a^2-b^2}\over {a^2}}.
\end{equation}
Here $K(k), E(k)$ and $\Pi(\nu^2,k)$ are complete elliptic integrals of 
first, second and third kind, respectively. Similarly we obtain for 
the third-order contribution to $\alpha$ (see eqs. (3.13b), (A.5b), (A.6b), 
(A3), (2.9b) and (3.2b) of \cite{fro1}),
\begin{eqnarray}
\alpha^{(3)} & = & {{1}\over {2}}\int_{\Lambda_\alpha}q^{(3)}(\eta) d\eta
\nonumber \\
& = & {{1}\over {2}} \int_{\Lambda_{\alpha}}\left[\left (-C+{{1}\over{1-\eta^2}}
\right ){{1}\over {2Q(1-\eta^2)}}
-{{1}\over{8}}
Q^{-3}(\eta)\left({{dQ(\eta)}\over{d\eta}}\right )^2\right]d\eta,
\end{eqnarray}
where $C$ is the parameter introduced in the base function $Q(\eta)$
in eq. (3.2b) of \cite{fro1}. After
evaluation of the integrals we obtain
\begin{align}
\alpha^{(3)} = & -{{gC}\over {2p}}K(k)+{{g}\over{2p(1-b^2)\nu^2}}
\left[k^2K(k)+(\nu^2-k^2)\Pi(\nu^2,k)\right]\nonumber\\
&-{{g(1-b^2)}\over{8pb^2k^4}}[P_1 K(k)+P_2 E(k)+ P_3 \Pi(\nu^2,k)],\tag{3.4b}
\end{align}
where
\begin{subequations}
\begin{align}
P_1 & = {{1}\over{3}}[-9k^4+k^2(8+5\nu^2)+4\nu^2-8], \\
P_2 & = {{1}\over{3}}[k^2(-4-\nu^2)+(8-4\nu^2)]
\end{align}
and
\begin{equation}
P_3=4(\nu^2-k^2)^2.
\end{equation}
\end{subequations}

Analogous calculations can be performed to evaluate the quantity
$\bar K$. For this purpose we make use of the appropriate transformation 
on p. 58 in \cite{byrd}, that is 
\begin{equation}
\label{b2}
\eta^2=b^2sn^2u.
\end{equation}
The first-order (see eqs.(3.15b), (A5b) and (A6a) of \cite{fro1})
and the third-order (see eqs. (3.15b), (A5b), (A6b), A(3), (2.9b) and (3.2b)
 of \cite{fro1})
contributions to $K$ $(=\pi\bar K)$ are
\begin{subequations}
\label{keq}
\begin{eqnarray}
\pi \bar K_0&=&\frac{i}{2}\int_{\Lambda_{K}}Q(\eta)d\eta \nonumber\\ 
&=&\frac{p\nu^2}{g}\int_0^{2K}{{cn^2udn^2u}\over {1-\nu^2sn^2u}}du\nonumber\\
&=&{{p}\over {g}}\left [E(k)+k^2
\left (1-{{1}\over {\nu^2}}\right )K(k)+(\nu^2-k^2)\left(1-{{1}\over {\nu^2}}\right ) 
\Pi(\nu^2,k)\right ],
\end{eqnarray}
and
\begin{eqnarray}
\pi\bar K_2&=&\frac{i}{2}\int_{\Lambda_{K}}q^{(3)}(\eta)d\eta\nonumber\\
&=&-2\mbox{Im}{{1}\over {2}}\int_{\Lambda_{K}}
\left [\left (-C+{{1}\over {1-\eta^2}}\right )
{{1}\over {2Q(1-\eta^2)}}-{{1}\over{8}}Q^{-3}(\eta)\left ({{dQ(\eta)}\over
{d\eta}}\right )^2\right ]d\eta\nonumber\\
&=& {{Cg}\over {p}}K(k)-{{g}\over {p}}\Pi(\nu^2,k)+{{g}\over {4b^2p}}\left[
{{1}\over {3k^{'6}}}(-3\nu^2k^6-k^4+8\nu^2k^2-7\nu^2+1)E(k)\right.\nonumber\\
&&\left.+{{3\nu^2-1}\over {3k^{'2}}}K(k)+4\nu^2\Pi(\nu^2,k)\right ],
\end{eqnarray}
\end{subequations}
where 
\begin{equation}
\nu^2=b^2, \hskip 10ptg={{1}\over {a}},\hskip 10pt k^2={{b^2}\over {a^2}},
\hskip 10pt k^{'2}=1-k^2=\frac{a^2-b^2}{a^2}.
\end{equation}

The integrals $\alpha '$ and $\beta '$ for the contours $\Lambda_{\alpha'}$
and $\Lambda_{\beta'}$ in Fig. 3(a) in \cite{fro1} are obtained from the formulas
$\alpha'=\alpha+\frac{\Lambda\pi}{2}$ and $\beta'=\beta+\frac{\Lambda\pi}{2}$;
see eq. (3.18a) in \cite{fro1}. 

\subsubsection{The quantities  $\tilde L $  and $\tilde L'$ pertaining 
to the $\xi$-equation [Fig. 1(a) in Ref. 1]}
\vskip 10 pt
Denoting the four real zeros $\xi_1 <\xi_2<1<\xi_3<\xi_4$ of $\tilde Q^2(\xi)$ by the simpler
notations $d<c<1<b<a$, respectively, used in \cite{byrd}, we have

\begin{equation}
\tilde Q(\xi)=p{{[(a-\xi)(\xi-b)(\xi-c)(\xi-d)]^{\frac {1}{2}}}\over 
{\xi^2-1}}.
\end{equation}
Using the appropriate transformation on p. 120 in \cite{byrd}, 
we obtain (cf. sec. II)
\begin{equation}
\label{tr}
\xi = {{b-c\nu_1^2sn^2u}\over {1-\nu_1^2sn^2u}},\hskip 10pt \nu_1^2={{a-b}\over
{a-c}}<1.
\end{equation}
\noindent Noting that the loop $\xi_3\rightarrow\xi_4\rightarrow\xi_3$, that is 
$b\rightarrow a\rightarrow b$, in the $\xi$-plane, denoted by $\Gamma_{b,a}$, represents
the contour $\Lambda_{\tilde L}$ in Fig. 1(a) of \cite{fro1} and
corresponds to $0\rightarrow K\rightarrow 2K$ in the $u$-plane, and using
the transformation (\ref{tr}), we obtain the first-order contribution to $\tilde L$
through the following transformation of the original integral in the $\xi$-plane
to the $u$-plane:
\vskip 10pt

\begin{eqnarray}
\tilde L^{(1)}&=&{{1}\over {2}}\int_{\Lambda_{\tilde L}}\tilde Q(\xi)d\xi\nonumber\\
&=&{{1}\over {2}}\int_{\Gamma_{b,a}}\tilde Q(\xi)d\xi\nonumber\\
&=&{{p}\over{2g}}\int_0^{2K}\left({{d\xi}\over {du}}\right)^2{{du}\over{\xi^2-1}}\nonumber\\
&=&{{2p(\nu_2^2-\nu_1^2)(\nu_3^2-\nu_1^2)}\over {g}}\int_0^{2K}
{{sn^2u(1-sn^2u)(1-k^2sn^2u)}\over {(1-\nu_1^2sn^2u)^2(1-\nu_2^2sn^2u)
(1-\nu_3^2sn^2u)}}du,\label{xi1}
\end{eqnarray}
where 
\begin{equation}
\nu_1^2={{a-b}\over{a-c}},\hskip 10pt \nu_2^2={{1+c}\over {1+b}}\nu_1^2,
 \hskip 10pt \nu_3^2={{1-c}\over {1-b}}\nu_1^2,
\end{equation}

\begin{equation}
g={{2}\over {[(a-c)(b-d)]^{\frac{1}{2}}}}, \hskip 10pt k^2={{(a-b)(c-d)}\over
{(a-c)(b-d)}}.
\end{equation}
Note that $a,b,c,d,$ and hence also $\nu_1,\nu_2,\nu_3,g,k$, depend on
the choice of the parameter $\tilde C$ in the base function $\tilde
Q(\xi)$; cf. (3.2a) in \cite{fro1}.
Decomposing the integrand in (\ref{xi1}) into partial fractions, and using recurrence
formulas in \cite{byrd}, we obtain the final formula
\begin{subequations}
\label{ll2}
\begin{eqnarray}
\tilde L^{(1)}&=&-H^{(1)}(\nu_1,\nu_2,\nu_3,g,k,\tilde C)\label{uni}\\
&=&{{2p}\over {g}}\left [\left(1-2k^2+{{3k^2}\over {\nu_1^2}}
\right )K(k) -3E(k)\right. \nonumber\\
&&\left.+ \left (2(1+k^2)-\nu_1^2-{{3k^2}\over {\nu_1^2}}\right )
\Pi(\nu_1^2,k) - \sum_{i=1}^3C_iS_i\right ],
\end{eqnarray}
\end{subequations}
where we have introduced the ``universal'' function $H^{(1)}$, and 
\begin{subequations}
\begin{align}
C_1 & ={{2[2\nu_2^2\nu_3^2-\nu_1^2\nu_2^2-\nu_1^2\nu_3^2]}\over
{(\nu_2^2-\nu_1^2)(\nu_1^2-\nu_3^2)}},\\
C_2 & ={{2\nu_2^2(\nu_3^2-\nu_1^2)}\over {(\nu_2^2-\nu_1^2)
(\nu_3^2-\nu_2^2)}},\\
C_3 & ={{2\nu_3^2(\nu_2^2-\nu_1^2)}\over {(\nu_3^2-\nu_1^2)(\nu_2^2-\nu_3^2)}},
\end{align}
\end{subequations}
\begin{eqnarray}
S_i&=&{{1}\over {3k^2}}\left [(\nu_i^2+2\nu_i^2k^2-3k^2)K(k)-(\nu_i^2
+\nu_i^2k^2-3k^2)E(k)\right.\nonumber\\
&&\left.+{{3k^2}\over {\nu_i^2}}(1-\nu_i^2)(k^2-\nu_i^2)
[\Pi(\nu_i^2,k)-K(k)]\right],\;\;\;i=1,2,3,
\end{eqnarray}
which can also be written as
\begin{align}
S_i = {{k^{'2}}\over {3}}\left [\nu_i^2{{K(k)-E(k)}\over {k^2}}
+{{(3-2\nu_i^2)E(k)}\over{k^{'2}}}-3\Pi\left ({{k^2-\nu_i^2}\over {1-\nu_i^2}},
k\right)\right],\;\;\;i=1,2,3,\tag{3.18$'$}
\end{align}
the last formula being valid if $\nu_i^2$ and $k^2$ fulfil the conditions stated
in section 117.03 in \cite{byrd}. 

Similarly we get for the third-order contribution to $\tilde L$:
\begin{eqnarray}
\tilde L^{(3)}&=&{{1}\over {2}}\int_{\Lambda_{\tilde L}} 
\left[\left (\tilde C+{{1}\over {\xi^2-1}}\right ){{1}
\over {2\tilde Q(\xi)(\xi^2-1)}}-{{1}\over {8}}
 \tilde Q^{-3}(\xi)\left ({{d\tilde Q}\over
{d\xi}}\right)^2\right]d\xi,\nonumber
\end{eqnarray}
that is
\begin{align}
\tilde L^{(3)} = & -H^{(3)}(\nu_1,\nu_2,\nu_3,g,k,\tilde C)\tag{3.16c}\\
= & -{{g}\over {64p(\nu_2^2-\nu_1^2)(\nu_3^2-\nu_1^2)}}
\left [\sum_{i=1}^4C'_i K(k)+\sum_{i=1,}^4 D_i E(k)\right]\nonumber\\
& -{{g}\over {4p}}\left [ \tilde C+{{\nu_1^2(\nu_2^2-\nu_3^2)}\over {4\nu_2^2
(\nu_1^2-\nu_3^2)}}+{{\nu_1^2(\nu_2^2-\nu_3^2)}\over
 {4\nu_3^2(\nu_2^2-\nu_1^2)}}\right ] K(k)\tag{3.16d}, 
\end{align}
where $H^{(3)}$ is another ``universal'' function, $\tilde C$ is the
parameter in the base function $\tilde Q(\xi)$ (cf. eq. (3.2a) in \cite{fro1}),
and
\begin{subequations}
\begin{align}
C'_1 & ={{4}\over {3k^{'2}}}\left [ k^4+2k^2-2+(2\nu_1^2+\nu_2^2+\nu_3^2)
(1-2k^2)+\{2\nu_1^2(\nu_2^2+\nu_3^2)+\nu_2^2\nu_3^2+\nu_1^4\}(4-3k^2)\right ]
\nonumber\\
& +{{4}\over {3k^{'2}}}\left [-\{2\nu_1^2\nu_2^2\nu_3^2+\nu_1^4
(\nu_2^2+\nu_3^2)\}
{{11-10k^2}\over {k^2}}+{{\nu_1^4\nu_2^2\nu_3^2}\over {k^4}}
(14-6k^2-7k^4)\right ],\\
C'_2 & ={{4}\over {\nu_2^2\nu_3^2}} \left [2\nu_2^4\nu_3^4+\nu_1^4\nu_2^4+
\nu_1^4\nu_3^4-2\nu_1^4\nu_2^2\nu_3^2-{{2\nu_1^2\nu_2^4
\nu_3^4}\over {k^2}}\left (2-{{\nu_1^2}\over {3}}\right )
+{{4\nu_1^4\nu_2^4\nu_3^4}\over {3k^4}}\right ],\\
C'_3 & ={{8\nu_2^2\nu_3^2}\over {3k^4}}(3k^4-6\nu_1^2k^2+2\nu_1^4+k^2\nu_1^4),\\
C'_4 & = 16 \left [-(\nu_1^2\nu_2^2+\nu_2^2\nu_3^2+\nu_3^2\nu_1^2)
-{{\nu_1^4\nu_2^2\nu_3^2}\over {k^4}}(2+k^2)+{{\nu_1^2}\over {k^2}}(\nu_1^2\nu_2^2+\nu_1^2\nu_3^2+4\nu_2^2\nu_3^2) 
\right ],
\end{align}
\end{subequations}
\begin{subequations}
\begin{align}
D_1& ={{4}\over {3k^{'4}}}\left [2-3k^2-3k^4+2k^6
-(2\nu_1^2+\nu_2^2+\nu_3^2)(1-4k^2+k^4)\right]\nonumber\\
& +{{4}\over {3k^{'4}}}\left [-\{2\nu_1^2(\nu_2^2+\nu_3^2)+\nu_2^2\nu_3^2
+\nu_1^4\}(1+k^2)\right ]\nonumber\\
& + {{4}\over {3k^{'4}}}\left [
{{(11-20k^2+11k^4)\{2\nu_1^2\nu_2^2\nu_3^2
+\nu_1^4(\nu_2^2+\nu_3^2)\}}\over {k^2}}-{{\nu_1^4\nu_2^2\nu_3^2}\over {k^4}}(14-13k^2-13k^4+14k^6) \right ], \\
D_2 & ={{16\nu_1^2\nu_2^2\nu_3^2}\over {3k^4}}[3k^2-\nu_1^2(1+k^2)],\\
D_3 & =D_2, \\
D_4 &= {{16\nu_1^2}\over {k^2}}\left [-(\nu_1^2\nu_2^2
+\nu_1^2\nu_3^2+4\nu_2^2\nu_3^2)+{{2\nu_1^2\nu_2^2\nu_3^2}\over {k^2}}(1+k^2)\right ].
\end{align}
\end{subequations}

The integral $\tilde L'$ for the contour $\Lambda_{\tilde L'}$ 
in Fig. 1(a) in \cite{fro1}
is obtained from the formula $\tilde L'=\tilde L+\frac{|m|}{2}\pi$.
Therefore $\tilde L^{'(1)}$ and $\tilde L^{'(3)}$ can be obtained 
from (3.16a,b) and (3.16c,d).

\subsection{Two real and two complex conjugate zeros of $Q^2(\eta)$ and
$\tilde Q^2(\xi)$ }

\subsubsection{The quantities $\alpha =\beta$, $\bar K $, $L$ and $L'$
pertaining to the $\eta$-equation: Superbarrier case [Fig. 4(a) or Fig.
2 in Ref. 1]} Putting $\eta_4=-\eta_1 =a$, $\eta_2 =\eta_3^*=-ia_1$, we
have

\begin{equation}
Q(\eta)=p{{[(a^2-\eta^2)(a_1^2+\eta^2)]^{\frac {1}{2}}}\over
{1-\eta^2}}.
\end{equation}
Using the appropriate transformation on p. 133 in \cite{byrd},
that is, 
\begin{equation}
\eta^2=a^2cn^2u,
\end{equation}
we now utilize the fact that the Jacobian elliptic functions are doubly
periodic, one of the periods being complex. Thus, 
the loop $\eta_1\rightarrow\eta_2\rightarrow\eta_1$, that is
$-a\rightarrow -ia_1\rightarrow -a$, in the $\eta $-plane, denoted by $\Gamma_{-a,-ia_1}$, represents the
contour  $\Lambda_\alpha$ in Fig. 4(a) of \cite{fro1}, and it corresponds in the $u$-plane to
$0\rightarrow K+iK' \rightarrow 2K+2iK'$ where $K'=K(k')$.
Denoting by $2\bar G^{(2n+1)}$ the integral occurring in the definitions (3.13b)
and $(3.15b')$ in \cite{fro1} when $Z_1=Z_2$, we have the following first-order 
expression:
\begin{eqnarray}
\label{xx1}
\bar G^{(1)}&=&{{1}\over {2}}\int_{\Lambda_{\alpha}}Q(\eta)d\eta\nonumber\\
&=&{{1}\over {2}}\int_{\Gamma_{-a,-ia_1}}Q(\eta) d\eta,
\end{eqnarray}
which when transformed to the variable $u$ becomes
\begin{equation}
\bar G^{(1)}= {{pa^2}\over {2g}}\int_0^{2K+2iK'} {{sn^2udn^2u}
\over {1-a^2cn^2u}}du.\label{xi2}
\end{equation}
After evaluation of the integral in (\ref{xi2}) we obtain 
\begin{subequations}
\label{super}
\begin{eqnarray}
\bar G^{(1)}&=&{{p}\over {g}}\left [E(k)+{{k^2}\over {a^2}}(1-a^2)K(k)
-\left ({{k^2}\over {a^2}}+k^{'2}\right )\Pi(\nu^2,k)\right ]\nonumber\\
&&-{{p}\over {g}}i\left\{ E(k')+(k^2+a^2k^{'2})\left[\Pi(1-\nu^2,k')-K(k')\right]\right\},
\end{eqnarray}
\end{subequations}
where
\begin{equation}
 \nu^2={{a^2}\over {a^2-1}},\mbox{ } g={{1}\over {(a^2+a_1^2)^{\frac{1}{2}}}}, \mbox{ }
k^2={{a^2}\over {a^2+a_1^2}},
\mbox{ }k^{'2}=1-k^2.
\end{equation}

Similarly one obtains
\begin{align}
\bar G^{(3)}& = -{{Cg}\over {2p}}K(k)
+\left \{{{g}\over {2p(1-a^2)}}+{{(\nu^2-1)g}\over {4p}}\right \}
\left [\Pi\left (\nu^2,k\right)-\frac{\pi}{2}
\sqrt{{{a^2(1-a^2)}\over{k^{'2}a^2+k^2}}}\right]\nonumber\\
&-{{g}\over{8\nu^2p}}\left[\frac{1}{3}(-4k^2+4\nu^2-1)K(k)\
+\frac{1}{3k^{'2}}
(-8k^4+8k^2(1-\nu^2)+1+7\nu^2)E(k)\right]\nonumber\\
&+i\left\{-{{Cg}\over {2p}}K(k')+\left[{{g}\over {2p(1-a^2)}}
+{{(\nu^2-1)g}\over {4p}}\right]
\left [(1-a^2)K(k')-\nu^2\Pi\left (\nu^2-1,k'\right)-\frac{\pi}{2}
\right]\right\}\nonumber\\
&-i{{g}\over{8\nu^2p}}\left[\frac{1}{3}(-4k^2+4\nu^2-1)K(k')
+\frac{1}{3k^{'2}}(-8k^4+8k^2(1-\nu^2)+1+7\nu^2)[K(k')-E(k')]\right].
\tag{3.25b}
\end{align}

Recalling the definition of $\bar G^{(2n+1)}$ above (\ref{xx1})
in the present paper, one sees that according to (3.13b) in \cite{fro1}
the first- and third-order contributions to $\alpha$ are
\begin{subequations}
\label{alp1}
\begin{equation}
\alpha ^{(1)}= \mbox{Re} \bar G^{(1)},
\end{equation}
\begin{equation}
\alpha^{(3)}=\mbox{Re} \bar G^{(3)},
\end{equation}
\end{subequations}
and that according to $(3.15b')$ in \cite{fro1} the first- and third-order
contributions to $K(=\pi\bar K)$
are
\begin{subequations}
\begin{equation}
\pi \bar K_0 =-2 \mbox{Im} \bar G^{(1)},
\end{equation}

\begin{equation}
\pi\bar K_{2}=-2 \mbox{Im} \bar G^{(3)}.
\end{equation}
\end{subequations}

The integrals $\alpha'=\beta'$ associated with the contours $\Lambda_{\alpha'}$
and $\Lambda_{\beta'}$ in Fig. 4 in \cite{fro1} are obtained from the integrals
$\alpha=\beta$
by means of the relations (3.18a) in \cite{fro1}, that is, $\alpha'=\beta'=
\alpha+\frac{\Lambda\pi}{2}=\beta+\frac{\Lambda\pi}{2}$. The integrals $L$ and
$L'$ associated with the contours $\Lambda_L$ and $\Lambda_{L'}$ in
Fig. 4(a) in \cite{fro1} can be obtained from the integrals $\alpha=\beta$ by means of
the formulas $L=\alpha+\beta=2\alpha$ and $L'=L+|m|$.

The only essential difference between $-Q^2(\eta)$ in Fig. 4(a) in \cite{fro1}
and $-Q^2(\eta)$ in Fig. 2 in \cite{fro1} is that in the former figure there
is an underdense barrier of $-Q^2(\eta)$, while in the latter figure 
$-Q^2(\eta)$ has a single minimum. For the case in Fig. 2 one has therefore
the formula $L^{(2n+1)}=\alpha^{(2n+1)}+\beta^{(2n+1)}=2\alpha^{(2n+1)}$,
with the expansions (3.27a) and (3.27b) for $\alpha^{(1)}$ and $\alpha^{(3)}$,
and the formula $L^{'(2n+1)}=L^{(2n+1)}+|m|\delta_{m,0}$. The case in Fig. 2
has, however not appeared in our applications.

\subsubsection{The quantities $\tilde L$ and $\tilde L'$ pertaining to
the $\xi$-equation [Fig. 1(a) in Ref. 1]} Denoting by $\xi_3=b$ and
$\xi_4=a$ the real zeros of $\tilde Q^2(\xi)$, and by $c$ and $c^{*}$
the complex conjugate zeros $\xi_1$ and $\xi_2$ of $\tilde Q^2(\xi)$,
we have

\begin{equation}
\tilde Q(\xi)=p{{[(a-\xi)(\xi-b)(\xi-c)(\xi-c^*)]^{\frac{1}{2}}}
\over {\xi^2-1}}.
\end{equation}
Defining
\begin{equation}
c=b_1-ia_1, \hskip 10pt c^*=b_1+ia_1,
\end{equation}

\begin{equation}
A=[(a-b_1)^2+a_1^2]^\frac{1}{2},\hskip 10pt B=[(b-b_1)^2+a_1^2]^\frac{1}{2},
\end{equation}
and using the appropriate transformation on p. 133 in \cite{byrd}, we get
\begin{equation}
\xi= {{aB+bA+(bA-aB)cnu}\over {A+B+(A-B)cnu}}. \label{xi3}
\end{equation}
Noting that the loop $b\rightarrow a\rightarrow b$ in the $\xi$-plane 
(denoted by $\Gamma_{b,a}$ for the contour $ \Lambda_{\tilde L}$)
corresponds to the path $0\rightarrow 2K\rightarrow 4K$ in the $u$-plane,
and using the transformation (\ref{xi3}), we obtain for the first-order contribution to
$\tilde L$ the formula (see Fig. 1(a) in \cite{fro1}),

\begin{eqnarray}
\tilde L^{(1)}&=&{{1}\over {2}}\int_{\Lambda _{\tilde L}}\tilde Q(\xi)d\xi\nonumber\\
&=&{{1}\over {2}}\int_{\Gamma_{b,a}} \tilde Q(\xi)d\xi\nonumber\\
&=&{{p(\nu_1-\nu_2)(\nu_1-\nu_3)}\over {g}}\int_0^{4K}
{{sn^2udn^2u}\over {(1+\nu_1cnu)^2(1+\nu_2cnu)(1+\nu_3cnu)}}du,\label{xi4}
\end{eqnarray}
where
\begin{equation}
\nu_1= {{A-B}\over {A+B}},\hskip 10pt \nu_2={{(1+b)A-(1+a)B}\over
{(1+b)A+(1+a)B}},\hskip 10pt
\nu_3={{(1-b)A-(1-a)B}\over {(1-b)A+(1-a)B}},
\end{equation}

\begin{equation}
g={{1}\over {\sqrt{AB}}}, \hskip 10pt k^2={{(a-b)^2-(A-B)^2}\over {4AB}}.
\end{equation}
By evaluating the last integral in (\ref{xi4}), and introducing a new ``universal''
function $\bar H^{(1)}$, given by eqs. (2.19), (2.20a-c) and (2.21) in
\cite{ath3} with $j=0$, we obtain
\begin{subequations}
\label{ll1}
\begin{eqnarray}
\tilde L^{(1)}&=&-2\mbox{Re}\bar H^{(1)}(\nu_1,\nu_2,\nu_3,g,k,\tilde C)\mbox{[with j=0]}\\
&=&{{2p}\over {g}}\left \{ {{1}\over {\nu_1^2}}\left [\left(2k^2+{{\nu_1^2}\over {1-\nu_1^2}}\right )
\Pi\left ({{\nu_1^2}\over {\nu_1^2-1}},k\right )
\right.\right.\nonumber\\
&&\left.\left. +(\nu_1^2-2k^2)K(k)-2\nu_1^2E(k)+\nu_1k\pi\right ]
+\sum_{i=1}^3 \bar C_i \bar J_i\right\},
\end{eqnarray}
\end{subequations}
where
\begin{subequations}
\begin{equation}
\bar C_1 = {{\nu_1^2(2\nu_3\nu_2-\nu_1\nu_3-\nu_1\nu_2)}\over
{(\nu_1-\nu_2)(\nu_1-\nu_3)}},
\end{equation}

\begin{equation}
\bar C_2= {{(\nu_1-\nu_3)\nu_2^3}\over {(\nu_1-\nu_2)(\nu_2-\nu_3)}},
\end{equation}

\begin{equation}
\bar C_3={{(\nu_1-\nu_2)\nu_3^3}\over {(\nu_1-\nu_3)(\nu_3-
\nu_2)}},
\end{equation}
\end{subequations}
\begin{eqnarray}
\bar J_i&=& \mbox{Re} \bar S_i(\mbox{with } j=0)\nonumber\\
&=&{{1}\over {\nu_i^4}}\left [k^2(1-\nu_i^2)K(k)+\nu_i^2E(k)
-(k^2+\nu_i^2k^{'2})\Pi\left({{\nu_i^2}\over {\nu_i^2-1}},k\right )
\right]\nonumber\\
&&+{{1}\over {2k\nu_i^3}}\left[k^2(\nu_i^2-1)-{{\nu_i^2}\over {2}}
\right ]\pi, i=1,2,3;
\end{eqnarray}
cf. for the definition of $\bar S_i$ eq. (2.21) in \cite{ath3}.
The third-order contribution to $\tilde L$ is
\begin{subequations}
\begin{eqnarray}
\tilde L^{(3)}&=&-2\mbox{Re}\bar H^{(3)}(\nu_1,\nu_2,\nu_3,g,k,\tilde C)\mbox{[with j=0]}\\
&=&-{{g}\over {8p}}\left \{\left (4\tilde C+{{\nu_1^2(\nu_2-\nu_3)^2}
\over {\nu_2 \nu_3(\nu_1-\nu_3)(\nu_2-\nu_1)}}\right ) K(k)\right.\nonumber\\
&&+\left.{{1}\over {(\nu_1-\nu_3)(\nu_1-\nu_2)}}\left[X K(k)
+Y E(k)\right ] \right \},
\end{eqnarray}
\end{subequations}
where $\bar H^{(3)}$ is another ``universal'' function, given by eqs. (2.22)
and (2.23a-c) in \cite{ath3} with $j=0$, and where thus

\begin{subequations}
\begin{align}
X &= -{{(1+4k^2)}\over {3}}+{{(3+4k^{'2})}\over {3}}(\nu_1^2+2\nu_1\nu_2+2\nu_1\nu_3+\nu_2\nu_3)
-{{k^{'2}}\over {3k^2}}(17-4k^2)
\nu_1^2\nu_2\nu_3\nonumber\\
&-2\nu_1(\nu_1+\nu_2+\nu_3)+2\nu_2\nu_3+\nu_1^2\left (
{{\nu_2}\over {\nu_3}}+{{\nu_3}\over {\nu_2}}\right ),\\
Y&= {{1}\over {3k^{'2}}}(1+8k^2-8k^4)+{{4}\over {3}}(2k^2-1)(\nu_1^2+2\nu_1
\nu_2+2\nu_1\nu_3+\nu_2\nu_3)\nonumber\\
&+{{\nu_1^2\nu_2\nu_3}\over
{3k^2}}(17-8k^2+8k^4).
\end{align}
\end{subequations}

The integral $\tilde L'$ associated with the contour $\Lambda_{\tilde L'}$
in Fig. 1(a) in \cite{fro1} is obtained from the formula 
$\tilde L'=\tilde L+\frac{|m|}{2}\pi$.
One has therefore the formulas $\tilde L^{'(1)}=\tilde L^{(1)}
+\frac{|m|}{2}\pi$ and
$\tilde L^{'(3)}=\tilde L^{(3)}$ with $\tilde L^{(1)}$ and $\tilde L^{(3)}$ 
given by (3.39a) and (3.39b).
\section{Case $\Lambda=0$}
\label{sec4}
\subsection{Two real zeros of $Q^2(\eta )$ and $\tilde Q^2(\xi )$}
\subsubsection{The quantities $\alpha= \beta $ and $\bar K$ pertaining to the  
$\eta$-equation: Subbarrier case [Fig. 3(b) in Ref. 1]}
Putting $\eta_3=-\eta_2=b$ as before, we have

\begin{equation}
Q(\eta ) = p\left [{{(\eta ^2 -b^2)}\over {1-\eta ^2}}\right]^{\frac {1}{2}}.
\end{equation}

\noindent Using (3.2) with $a=1$, the first- and third-order contributions
to $\alpha$ become 
\begin{subequations}
\label{alp3}
\begin{equation}
\alpha^{(1)} =   p[E(k)-(1-k^2)K(k)],
\end{equation}  
\begin{eqnarray}
\alpha^{(3)}&=& {{1}\over {2pk^2}} \left [{(1-Ck^2) K(k)} -E(k)
-{{1}\over {12}} \left ((3k^2-8)K(k)
-(7k^2-8){{E(k)}\over {k^{'2}}}\right )\right ],
\end{eqnarray}
\end{subequations}
\noindent where $k^2=1-b^2$.
\vskip 5pt
The first- and third-order contributions to $K(=\pi\bar K)$ are

\begin{subequations}
\label{keq1}
\begin{equation}
\pi \bar K_0 =  2p[E(k)-(1-k^2)K(k)],
\end{equation}  
\begin{eqnarray}
\pi \bar K_2 &=&{{1}\over {p}}\left [CK(k)-{{E(k)}\over {k^{'2}}}+{{1}\over 
{12k^2}}\left ({{7k^2+1}\over {k^{'2}}}E(k)-(3k^2+1)K(k)\right ) \right ],
\end{eqnarray}
\end{subequations}
\noindent where $k^2=b^2$.

\subsubsection{The quantities  $\tilde L$ and $\tilde L'$ pertaining to
the $\xi$-equation [Fig. 1 in Ref. 1]}

If the real zeros of $\tilde Q^2(\xi )$ are $\xi_3=c(<a)$ and $\xi_4=a$, we
have

\begin{equation}
\tilde Q(\xi )=p {\left [ {{(\xi -c)(a-\xi )}\over {(\xi+1)(\xi -1)}} \right ]}^{\frac {1}{2}}.
\end{equation}
We shall treat the three cases $1<c<a$, $-1<c<1<a$ and
$c<-1<1<a$ separately.
\vskip 10pt
\noindent \underline {Case $1<c<a$} [Fig. 1(a) in Ref. 1]
\vskip 10pt
For this case we use in \cite{byrd} the transformation on p. 120 and 
the formula in section 256.19 with
a suitable choice of parameters  to obtain
the first- and third-order contributions to $\tilde L$ as
\begin{subequations}
\label{l10}
\begin{align}
\tilde L^{(1)}&=  \int_c^a \tilde Q(\xi )d\xi\nonumber\\
&= p{{(a-c)(c-1)g}\over {2\nu^2(\nu^2-k^2)}}\left [(2\nu^2-\nu^4
-k^2)\Pi(\nu^2,k)-\nu^2E(k)-(\nu^2-k^2)K(k)\right ],\\
\tilde L^{(3)}&= {{\tilde CgK(k)}\over {2p}}+{{g}\over {2(c^2-1)p}}\left [{{2\nu^2}
\over{k^4}}(k^2-\nu^2)K(k)+{{1}\over {k^4k^{'2}}}\left (k^2(k^2-\nu^4-2\nu^2)
+2\nu^4\right )E(k)\right ]\nonumber\\
& -{{g}\over {24\nu^4p(c-1)}}\left [ (-2+k^2+2\nu^2-\nu^4)K(k)
+2\left (k^4-(1+\nu^2+\nu^4)k^2-\nu^2+2\nu^4+1\right )E(k)\right ],
\end{align}
\end{subequations}
where 
\begin{equation}
\nu^2={{a-c}\over {a-1}}, \hskip 5pt g={{2}\over {[(a-1)(c+1)]}^{\frac {1}{2}}},\hskip 5pt
 k^2={{2(a-c)}\over {(a-1)(c+1)}}.
\end{equation}

To obtain $\tilde L'$ one can use the formula $\tilde L^{'(2n+1)}=\tilde L^{(2n+1)}+
\frac{|m|}{2}\delta_{m,0}$ with $\tilde L^{(1)}$ and $\tilde L^{(3)}$ given by (4.5a)
and (4.5b).

\noindent \underline {Case $-1<c<1<a$} [Fig. 1(b) in Ref. 1]
\vskip 10pt
We use in \cite{byrd} the appropriate transformation on p. 120 and the formula
in section 256.17 (with $b=1$ and $d=-1$) 
to obtain the first- and third-order
contributions to $\tilde L'$ (see Fig.1(b) in \cite{fro1})
as
\begin{subequations}
\label{l11}
\begin{eqnarray}
\tilde L^{'(1)}&=&\frac{2p}{g}\left [\left (1-\frac{k^2}{\nu^2}\right )K(k)
-E(k)+\left (\nu^2-2k^2+\frac{k^2}{\nu^2}\right )\Pi(\nu^2,k)\right ],
\end{eqnarray}
\begin{eqnarray}
\tilde L^{'(3)}&=&{{g\tilde CK(k)}\over {2p}}+{{g}\over {4p\nu^2(1-c)}}\left [
\left(1-{{\nu^4}\over {k^2}}\right )K(k)+\left (k^2(2k^2-1-2\nu^2)+\nu^4\right )
{{E(k)}\over {k^2k^{'2}}} \right ]\nonumber\\
&&-{{g}\over {16(1-c)p\nu^2}}\left [{{4}\over {3k^2}}\left ( (2+\nu^4+2\nu^2)
k^2-3k^4-2\nu^4\right )K(k)\right.\nonumber\\
&&\left.+{{8}\over {3k^{'2}}}
\left (-(\nu^4+\nu^2-1)+k^2(\nu^4-\nu^2+2)+{{\nu^4}\over {k^2}}\right )E(k)\right ],
\end{eqnarray}
\end{subequations}
where 
\begin{equation}
\nu^2={{a-1}\over {a-c}},\hskip 5pt g= \left ({{2}\over {a-c}}
\right )^{\frac{1}{2}},\hskip 5pt k^2={{(1+c)(a-1)}\over {2(a-c)}}.
\end{equation}

\underline{Case $c<-1<1<a$} [Fig. 1(b) in Ref. 1]

Using in \cite{byrd} the transformation on p. 120 and the formula in section 256.20
 with a suitable choice of parameters,
we obtain the first- and third-order contributions to $\tilde L'$ as
\begin{subequations}
\begin{eqnarray}
\tilde L^{'95(1)}&=&{{(a-c)(1-c)gp}\over{2\nu^4}}\left[-\nu^2E(k)+(\nu^2+
k^2)K(k)+(\nu^4-k^2)\Pi(\nu^2,k)\right],
\end{eqnarray}
\begin{eqnarray}
\tilde L^{'(3)}&=&{{\tilde CgK(k)}\over {p}}+{{g}\over{4p\nu^2}}\left[(1-2\nu^2+{{\nu^4}\over {k^2}})K(k)
-(1+{{\nu^4}\over {k^2}})E(k)\right]\nonumber\\
&&-{{g}\over{32p\nu^2}}\left\{{{4}\over{3k^2k^{'2}}}\left[-k^4+k^2(2-\nu^4-2\nu^2)-\nu^4
\right]K(k)\right.\nonumber\\
&&\left.+{{8}\over{3k^2k^{'4}}}\left[-k^6+k^4(\nu^2-\nu^4)+k^2(\nu^4+\nu^2-1)-\nu^4\right]E(k) \right \},
\end{eqnarray}
\end{subequations}
with 
\begin{equation}
\nu^2={{a-1}\over {a+1}},\hskip 5pt g={{2}\over {[(a+1)(1-c)]^\frac{1}{2}}},\hskip 5pt 
k^2={{(a-1)(-c-1)}\over{(a+1)(1-c)}}.
\end{equation}

\subsection{Two complex conjugate zeros of $Q^2(\eta)$}

The case of two complex conjugate zeros of the square of the base
function occurs only for the $\eta$-equation.

\subsubsection{ The quantities $\alpha =\beta$ and $\bar K$ pertaining
to the $\eta $-equation: Superbarrier case \mbox{ } [Fig. 4(b) in Ref.
1]}

With $\eta_2=-ia_1$ and $\eta_3=ia_1$ we have

\begin{equation}
Q(\eta)=p{{[(1-\eta^2)(a_1^2+\eta^2)]^{\frac{1}{2}}}\over {1-\eta^2}}
=p\left({{a_1^2+\eta^2}\over {1-\eta^2}}\right )^{\frac{1}{2}}.
\end{equation}

\noindent Specializing to the case $a=1$ in (3.27a,b) and (3.28a,b) along with (3.25a,b), 
we obtain the first- and third-order contributions to $\alpha$ as
\begin{equation}
\label{r120}
\alpha^{(1)}=\mbox{Re} \bar G^{(1)},\hskip 5pt \alpha^{(3)}=\mbox{Re} \bar G^{(3)}
\end{equation}
and the first- and third-order contributions to $K(=\pi\bar K)$ as
\begin{equation}
\pi \bar K_0 =-2\mbox{Im} \bar G^{(1)},\hskip 5pt \pi \bar K_2 =-2\mbox{Im} \bar G^{(3)},
\end{equation}

where
\begin{subequations}
\begin{equation}
\bar G^{(1)}={{p}\over {g}}\left \{E(k)+i[K(k')-E(k')\right ]\},
\end{equation}
\begin{eqnarray}
\bar G^{(3)}&=&{{g}\over {2p}}\left [(1-C)K(k)-E(k)
-{{1}\over {12}}\left (
4K(k)+{{(8k^2-7)}\over {k^{'2}}}E(k)\right )\right ]\nonumber\\
&&+i{{g}\over {2p}}\left [-CK(k')+E(k')-{{1}\over {12}}
\left (4K(k')+{{(8k^2-7)}\over {k^{'2}}}\{
K(k')-E(k')\}\right ) \right] 
\end{eqnarray}
\end{subequations}
with
\begin{equation}
g={{1}\over {(1+a_1^2)^{\frac{1}{2}}}},\hskip 10pt k^2={{1}\over {1+a_1^2}}.
\end{equation}

The integral $L'$ associated with the contour $\Lambda_{L'}$ in Fig. 4(b) 
in \cite{fro1} is obtained from the formula $L'=\alpha +\beta =2\alpha$.

\section{accuracy of the phase-integral quantization conditions for the
$1\lowercase{s}\sigma$ and $2\lowercase{p}\sigma$ states of the
hydrogen molecule ion} 
\label{sec5}

For the $1s\sigma$ and $2p\sigma$ states of the hydrogen molecule ion
one has to put $\Lambda=0$.  The quantization conditions in \cite{fro1}
for the $1s\sigma$ state are (3.5a) with $\tilde s=0$ [Fig. 1(a) in
Ref. 1] and (3.9) with $s=m=0$ [Fig. 4(b) in Ref. 1] when $r_{12}$ is
sufficiently small, but (3.5b) with $\tilde s=m=0$ [Fig. 1(b) in Ref.
1] and (3.25b) with $s_{\alpha}=s_{\beta}=m=0$ [Fig. 3(b) in Ref. 1]
when $r_{12}$ is sufficiently large.  The quantization conditions in
\cite{fro1} for the $2p\sigma$ state are (3.5a) with $\tilde s=0$ [Fig.
1(a) in Ref. 1] and (3.9) with $s=1$ and $m=0$ [Fig. 4(b) in Ref. 1]
when $r_{12}$ is sufficiently small, but (3.5b) with $\tilde s=m=0$
[Fig. 1(b) in Ref. 1] and (3.25a) with $s_{\alpha}=s_{\beta}=m=0$ [Fig.
3(b) in Ref. 1] when $r_{12}$ is sufficiently large.  After having
expressed these quantization conditions in the first and third order of
the phase-integral approximation in terms of complete elliptic
integrals, as described in the previous sections, we have used these
quantization conditions to calculate the energy $E$ and the reduced
separation constant $A'$.  In subsection \ref{sec:5:1} we determine $C$ and
$\tilde C$ as functions of $r_{12}$ such that the first- and
third-order quantization conditions give the same values of both $E$
and $A'$. For the values of $C$ and $\tilde C$ thus obtained, the
choice of the base functions $Q(\eta)$ and $\tilde Q(\xi)$ is optimal
in the sense that the most accurate first-order values of $E$ and $A'$
are obtained, since the first- and third-order approximations give the
same values of $E$ and $A'$. In this connection we remark that there
are quantal systems for which one can obtain exact values of the energy
by choosing the base function such that the first- and third-order
results coincide; see p. 1826 in \cite{fro2} and p. 16 in \cite{fro3}.
In subsection \ref{sec:5:2} we determine $C$ and $\tilde C$ such that the
phase-integral quantization conditions give the numerically exact
values of $E$ and $A'$ obtained by Murai and Takatsu \cite{Murai,Mu1}
and establish that the values of $C$ and $\tilde C$ thus obtained are
in qualitative agreement with the values of $C$ and $\tilde C$
determined in subsection \ref{sec:5:1}.

\subsection{Determination of $C(r_{12})$ and $\tilde C(r_{12})$ such
that the first- and third-order quantization conditions give the same
results}
\label{sec:5:1}

By determining $C$ and $\tilde C$ for each value of $r_{12}$ such that
the first- and third-order quantization conditions give the same value
of $E$ as well as of $A'$, we have obtained the results in Table I for
the $1s\sigma$ state and in Table II for the $2p\sigma$ state of the
hydrogen molecule ion;  see also Figs. 1 and 2. The phase-integral
values of $E$ and $A'$ in these tables are in reasonable agreement with
the numerically exact values obtained by Murai and Takatsu
\cite{Murai,Mu1}, as is best seen from Fig.1 for the state $1s\sigma$
and from Fig. 2 for the state $2p\sigma$.

It is seen that in Figs. 1 and 2 there are sometimes some
irregularities  in the values of $C$ , $\tilde C$, $|E-E_{MT}|$ and
$|A'-A'_{MT}|$ for low $r_{12}$ values.  It should also be noted that
$C$ and $\tilde C$ approach the correct limiting value $\frac{1}{4}$
as  $r_{12}\rightarrow 0$.

\subsection{Determination of $C(r_{12})$ and $\tilde C(r_{12})$ such
that the first-order phase-integral quantization conditions reproduce
numerically exact values of $E$ and $A'$}
\label{sec:5:2}

By determining $C$ and $\tilde C$ for each value of $r_{12}$ such that
the first-order quantization conditions reproduce the numerically exact
values of $E$ and $A'$ calculated by Murai and Takatsu \cite{Murai,Mu1}
we have obtained the values of $C$ and $\tilde C$ presented in Table
III for the $1s\sigma$ state and in Table IV for the $2p\sigma$
state. In Figs. 3 and 4, we have shown the dependence of $C$
and $\tilde C$ on $r_{12}$ for the states $1s\sigma$ and $2p\sigma$,
respectively. One can carry out a similar calculation using the 
third-order phase-integral quantization condition also and determine the
appropriate $C$ and $\tilde C$. However, the numerical analysis becomes
too laborious and time consuming and so we have not presented the
results here.

To obtain the numerical results in subsections V.A and V.B a general 
FORTRAN computer program using very rapid library routines was written at the
Centre for Nonlinear Dynamics, Department of Physics, Bharathidasan
University, Tiruchirapalli, India.  We have carried out the numerical
calculations by Silicon Graphics Power Indigo 2 XZ Graphics Workstation
(R8000, 64bit processor) using FORTRAN 77 compiler.

Some years ago, a direct numerical integration of the contour integrals
in the phase-integral quantization conditions for the hydrogen molecule
ion was carried out by Fil. lic. Anders H\"okback at the Department of
Theoretical Physics, University of Uppsala, Sweden. By means of this
numerical material it was possible to make valuable checks of the
correctness of the phase-integral quantization conditions expressed in
terms of complete elliptic integrals.

\acknowledgments{The authors are much indebted to Fil. lic. Anders
H\"{o}kback for placing his unpublished numerical material at their
disposal. The authors are extremely grateful to Professor Per Olof
Fr\"oman for very critical reading of the manuscript and for making
numerous comments which resulted in a much improved presentation.  The
work of M.L. forms part of a Department of Science and Technology,
Government of India, research project. Support from the Swedish Natural
Science Research Council for M. Lakshmanan's visits to Uppsala is
gratefully acknowledged.  }

\newpage
\input{table1}
\input{table2}
\input{table3}
\input{table4}

\begin{figure}
\centerline{\epsfig{figure=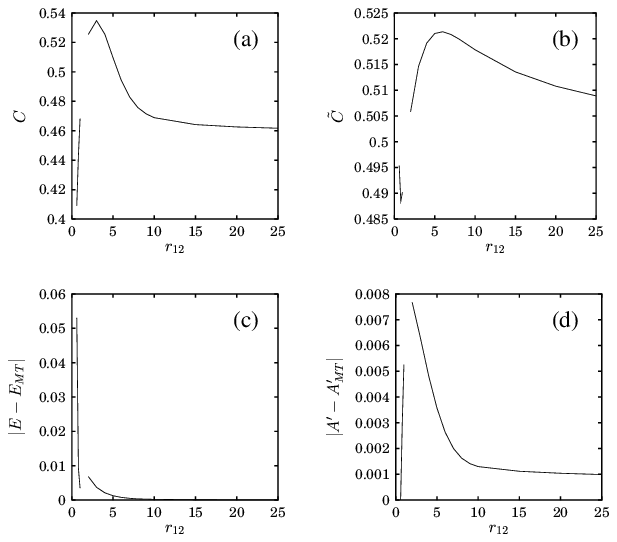, width=\linewidth}}

\caption {Plots for the $1s\sigma$ state of the ion $H_2^+$ of (a) $C$
versus $r_{12}$, (b) $\tilde C$ versus $r_{12}$, (c) $|E-E_{MT}|$
versus $r_{12}$ and (d) $A'-A'_{MT}$ versus $r_{12}$, when $C$ and
$\tilde C$ are determined as functions of $r_{12}$ from the requirement
that the first-order phase-integral results coincide with the third-
order results. Here $E$ and $A'$ are the phase-integral values obtained
in Table I, while $E_{MT}$ and $A'_{MT}$ are the corresponding
numerically exact values obtained by Murai and Takatsu [13,14] and
quoted in the same table.  There is a break in each curve between the
regions where the quantization conditions for sufficiently small and
sufficiently large values of $r_{12}$ have been used.}

\label{Fig.2(a)}
\end{figure}

\begin{figure}
\centerline{\epsfig{figure=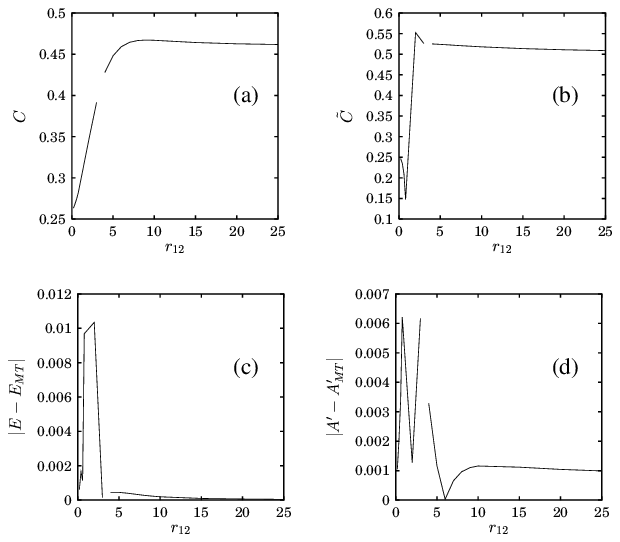, width=\linewidth}}

\caption{Plots for the $2p\sigma$ state of the ion $H_2^+$ of (a) $C$
versus $r_{12}$, (b) $\tilde C$ versus $r_{12}$, (c) $|E-E_{MT}|$
versus $r_{12}$ and (d) $A'-A'_{MT}$ versus $r_{12}$, when $C$ and
$\tilde C$ are determined as functions of $r_{12}$ from the requirement
that the first-order phase-integral results coincide with the third-
order results. Here $E$ and $A'$ are the phase-integral values obtained
in Table II, while $E_{MT}$ and $A'_{MT}$ are the corresponding
numerically exact values obtained by Murai and Takatsu [13,14] and
quoted in the same table.  There is a break in each curve between the
regions where the quantization conditions for sufficiently small and
sufficiently large values of $r_{12}$ have been used.}

\label{Fig.2(b)}
\end{figure}

\begin{figure}
\centerline{\epsfig{figure=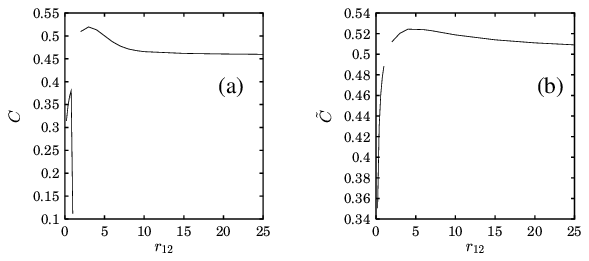, width=\linewidth}}

\caption{ Plots for the $1s\sigma$ state of the ion $H_2^+$ of
 (a) $C$ versus $r_{12}$ and (b) $\tilde C$  versus $r_{12}$, when $C$
and $\tilde C$ are determined as functions of $r_{12}$ from the
requirement that the first-order phase-integral results coincide with
the numerically exact results obtained by Murai and Takatsu [13,14].
There is a break in each curve between the regions where the
quantization conditions for sufficiently small and sufficiently large
values of $r_{12}$ have been used.}

\label{Fig.2(c)}
\end{figure}

\begin{figure}
\centerline{\epsfig{figure=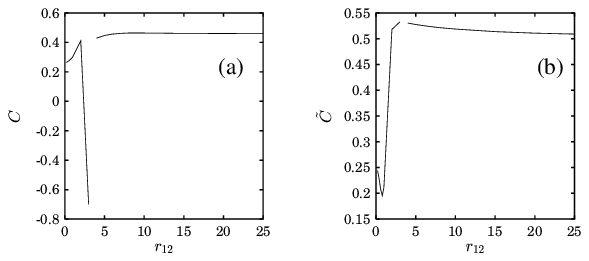, width=\linewidth}}

\caption{ Plots for the $2p\sigma$ state of the ion $H_2^+$ of (a) $C$
versus $r_{12}$ and (b) $\tilde C$  versus $r_{12}$, when $C$ and
$\tilde C$ are determined as functions of $r_{12}$ from the requirement
that the first-order phase-integral results coincide with the
numerically exact results obtained by Murai and Takatsu [13,14].  There
is a break in each curve between the regions where the quantization
conditions for sufficiently small and sufficiently large values of
$r_{12}$ have been used.}

\label{Fig.2(d)}
\end{figure}

\end{document}

%% file: table1.tex
\begin{table}[ht!]
\caption{
For the state $1s\sigma$ of $H_2^+$ the values of $C$ and $\tilde C$ have been obtained
from the requirement that the first- and third-order phase-integral results
coincide for $E$ as well as for $A'$. With the use of these values of $C$ and $\tilde C$,
the  values of $E$ and $A'$ have then been obtained from the 
quantization conditions that are appropriate depending
on whether $r_{12}$ is sufficiently small or sufficiently large. The numerically exact values
obtained by Murai and Takatsu [13,14] are given in the columns 
called $E_{MT}$ and $A'_{MT}$.
}
\begin{tabular}{c|cc|ccc|ccc}
\hline
$r_{12}$ & $C$ & $\tilde C$ & $E$ & $E_{MT}$ & $E-E_{MT}$ & $A'$&$A'_{MT}$ & $A'-A'_{MT}$ \\
\hline
\multicolumn{9}{l}{}\\ 
\multicolumn{9}{l} {Sufficiently small $r_{12}$}\\
\hline
0.6&0.4091767180&0.4953643560&-1.618424439&-1.6714847145&0.053060275&-0.1992407489&-0.1992300000&-0.000010748\\
\hline

0.8&0.4431528990&0.4884579844& -1.545199993&-1.5544800915&0.009280098&-0.3249142019&-0.327900000&0.002985799\\
\hline

1.0&0.4681771744&0.4902081420&-1.455314042&-1.4517863130&-0.003527729&-0.4706879721&-0.4759469161&0.005258944\\
\hline
\multicolumn{9}{l}{}\\ 
\multicolumn{9}{l} {Sufficiently large $r_{12}$}\\
\hline
2.0&0.5255080000&0.5057903368&-1.109450173&-1.1026342150&-0.006815958&-1.385866329&-1.393538844&0.007672515\\
\hline

3.0&0.5347955900&0.5146860140&-0.9146268870&-0.9108961974&-0.00373069&-2.451736492&-2.458030452&0.00629396\\
\hline

4.0&0.5255436520&0.5191506880&-0.7981985544&-0.7960848837&-0.002113671&-3.564275870&-3.569090310&0.00481444\\
\hline

5.0&0.5095344153&0.5210276591&-0.7256775451&-0.7244202952&-0.00125725&-4.673981394&-4.677559936&0.003578542\\ 
\hline

6.0 &0.4942331566&0.5213617485&-0.6794162838&-0.6786357151&-0.000780568&-5.759202308&-5.761839130&0.002636822\\ 
\hline

7.0 &0.4829917740&0.5208505340&-0.6489619089&-0.6484511471&-0.000510761&-6.817236218&-6.819239945&0.002003727\\   
\hline

8.0 &0.4758245700&0.5199462800&-0.6279265372&-0.6275703886&-0.000356149&-7.854452399&-7.856077820&0.001625421\\ 
\hline

9.0 &0.4715076730&0.5189057600&-0.6125711074&-0.6123065640&-0.000264543&-8.878337258&-8.879752233&0.001414975\\ 
\hline
 
10.0 &0.4688953184&0.5178568616&-0.6007859807&-0.6005787289&-0.000207251&-9.894343735&-9.895643269&0.001299534\\ 
\hline

15.0 &0.4641710460&0.5135686450&-0.5668106207&-0.5667156052&-0.000095015&-14.93203345&-14.93315205&0.00111865\\   
\hline

20.0&0.4625834317&0.5107892808&-0.5500740452&-0.5500142593&-0.000059786&-19.94891764&-19.94996067&0.00104303\\ 
\hline

25.0&0.4617350800&0.5089200200&-0.5400488855&-0.5400058008&-0.000043085&-24.95899430&-24.95998443&0.00099013\\
\hline
\end{tabular}
\end{table}

%% file: table2.tex
\begin{table}[ht!]
\caption{
For the state $2p\sigma$ of $H_2^+$ the values of $C$ and $\tilde C$
have been obtained from the requirement that the first- and third-order
phase-integral results coincide for $E$ as well as for $A'$. With the
use of these values of $C$ and $\tilde C$, the  values of $E$ and $A'$
have then been obtained from the quantization conditions that are
appropriate depending on whether $r_{12}$ is sufficiently small or
sufficiently large. The numerically exact values obtained by Murai and
Takatsu [13,14] are given in the columns called $E_{MT}$ and
$A'_{MT}$.
}
\begin{tabular}{c|cc|ccc|ccc}
\hline
$r_{12}$ & $C$ & $\tilde C$ & $E$ & $E_{MT}$ & $E-E_{MT}$ & $A'$&$A'_{MT}$ & $A'-A'_{MT}$ \\
\hline
\multicolumn{9}{l}{}\\ 
\multicolumn{9}{l} {Sufficiently small $r_{12}$}\\
\hline
0.2&0.2633959780&0.246516819&-0.5020720303&-0.5026800000&-0.00060797&-2.005074078&-2.004020000&0.001054078\\
\hline

0.4&0.2688550280&0.234829974&-0.5091302928&-0.5107900000&-0.001659708&-2.018393631&-2.016330000&0.002063631\\
\hline

0.6&0.2750685470&0.209525191&-0.5231745581&-0.5243050000&-0.001130442&-2.041072959&-2.037690000&0.003382959\\
\hline

0.8&0.2826340310&0.1470935880&-0.5524257935&-0.5427400000&0.009685793&-2.075482537&-2.069270000&0.006212537\\
\hline

2.0&0.3435728404&0.5524319489&-0.6571837692&-0.6675343922&-0.010350623&-2.523233873&-2.521958177&0.001275696\\
\hline

3.0&0.3915235980&0.5260048459&-0.7015678961&-0.7014183334&0.000149563&-3.202554526&-3.196382289&0.006172237\\
\hline
\multicolumn{9}{l}{}\\
\multicolumn{9}{l} {Sufficiently large $r_{12}$}\\
\hline
4.0& 0.4279941950&0.5249345450&-0.6960011808&-0.6955506394&0.000450542&-4.029227017&-4.025940635&0.003286382\\ 
\hline

5.0&0.4481616320&0.5238924640&-0.6777410414&-0.6772916132&0.000449428&-4.942443752&-4.941274459&0.001169293\\
\hline
 
6.0&0.4591233730&0.5226760960&-0.6577104664&-0.6573105590&0.000399907&-5.903631412&-5.903659889&-0.000028477\\
\hline

7.0&0.4644579210&0.5214162480&-0.6394674183&-0.6391288554&0.000338563&-6.890343283&-6.890997919&-0.000654636\\
\hline

8.0&0.4666144860&0.5201789140&-0.6238870346&-0.6236060156&0.000280969&-7.889744744&-7.890707161&-0.000962417\\
\hline

9.0&0.4671264910&0.5189985500&-0.6108876012&-0.6106549406&0.00023266&-8.894778515&-8.895880333&-0.001101818\\
\hline

10.0&0.4668770890&0.5178931230&-0.6000950569&-0.5999010686&0.000193989&-9.901798495&-9.902954530&-0.001156035\\
\hline

15.0&0.4641395400&0.5135689400&-0.5668036046&-0.5667087290&0.000094875&-14.93214472&-14.93326111&-0.00111639\\
\hline

20.0&0.4625830500&0.5107892800&-0.5500739820&-0.5500141977&0.000059785& -19.94891896&-19.94996191&-0.00104295 \\
\hline

25.0&0.4617350700&0.5089200100&-0.5400488850&-0.5400058003&0.000043085&-24.95899432&-24.95998445&-0.00099013\\
\hline
\end{tabular}
\end{table}

%% file: table3.tex
\begin{table}
\caption{
For the state $1s\sigma$ of $H_2^+$ the values of $C$ and $\tilde C$
have been obtained from the requirement that the first-order
phase-integral results, obtained from quantization conditions that are
appropriate depending on whether $r_{12}$ is sufficiently small or
sufficiently large, coincide for $E$ as well as for $A'$ with the
numerically exact results obtained by Murai and Takatsu [13,14] and
quoted  in this table as $E_{MT}$ and $A'_{MT}$.}
\begin{tabular}{ccccc}
\hline
$r_{12}$& $E_{MT}$  &   $ A'_{MT}$&  $C$ & $\tilde C$  \\
\hline
\multicolumn{5}{l} {}\\
\multicolumn{5}{l} {Sufficiently small $r_{12}$}\\
\hline
0.2&      -1.9286203017&    -0.0256900000& 0.3148379823&     0.3506512286 \\
\hline
0.4&     -1.8007540595&     -0.0957300000&  0.3477075009&     0.4296524070\\
\hline
0.6&       -1.6714847145&   -0.1992300000& 0.3681917193&     0.4607730539\\
\hline
0.8&       -1.5544800915&   -0.3279000000& 0.3807005347&     0.4777532944\\    
\hline
1.0&       -1.4517863130&   -0.4759469161& 0.1120058693&     0.4885447845\\    
\hline
\multicolumn{5}{l} {}\\
\multicolumn{5}{l} {Sufficiently large $r_{12}$}\\
\hline
2.0&      -1.1026342150&   -1.393538844& 0.5095379703&     0.5120864521  \\
\hline
3.0&       -0.9108961974&   -2.458030452&  0.5196342600&     0.5202577750\\   
\hline
4.0&       -0.7960848837&   -3.569090310&  0.5131695740&     0.5235297064\\  
\hline
5.0&       -0.7244202952&   -4.677559936&  0.5001127420&     0.5243749920\\  
\hline
6.0&       -0.6786357151&   -5.761839130&  0.4872379220&     0.5239023990\\ 
\hline
7.0&       -0.6484511471&   -6.819239945&  0.4776859510&     0.5227893710\\
\hline
8.0&       -0.6275703886&   -7.856077820&  0.4715798270&     0.5214462830\\  
\hline
9.0&       -0.6123065640&   -8.879752233&  0.4679041020&     0.5200867890\\   
\hline
10.0&      -0.6005787289&   -9.895643269&  0.4656852040&     0.5188037440\\  
\hline
15.0&      -0.5667156052&   -14.93315205&  0.4617106320&     0.5139609640\\  
\hline
20.0&      -0.5500142593&   -19.94996067&  0.4603989670&     0.5109975070\\  
\hline
25.0&      -0.5400058008&   -24.95998443&  0.4597077000&     0.5090477000\\
\hline
\end{tabular}
\end{table}

%% file: table4.tex
\begin{table}
\caption{
For the state $2p\sigma$ of $H_2^+$ the values of $C$ and $\tilde C$
have been obtained from the requirement that the first-order
phase-integral results, obtained from quantization conditions that are
appropriate depending on whether $r_{12}$ is sufficiently small or
sufficiently large, coincide for $E$ as well as for $A'$ with the
numerically exact results obtained by Murai and Takatsu [13,14] and
quoted  in this table as $E_{MT}$ and $A'_{MT}$.}
\begin{tabular}{ccccc}
\hline
$r_{12}$&  $E_{MT}$  &   $ A'_{MT}$& $C$ & $\tilde C$ \\
\hline
\multicolumn{5}{l} {}\\
\multicolumn{5}{l} {Sufficiently small $r_{12}$}\\
\hline
0.2&-0.5026800000&-2.004020000&0.264492583&0.2439809316\\
\hline
0.4&       -0.5107900000&    -2.016330000& 0.2712715280&    0.2275797176\\     
\hline
0.6&        -0.5243050000&    -2.037690000&0.2795333746&    0.2071325406\\     
\hline
0.8&       -0.5427400000&    -2.069270000&0.2897524086&    0.1952877316 \\     
\hline
1.0&       -0.5648136251&    -2.112417232& 0.3022903142&    0.2116506664\\    
\hline
2.0&       -0.6675343922&    -2.521958177& 0.4118503542&    0.5180809640 \\    
\hline
3.0&       -0.7014183334&    -3.196382289&-0.7005640546&    0.5327867225\\    
\hline
\multicolumn{5}{l} {}\\
\multicolumn{5}{l} {Sufficiently large $r_{12}$}\\
\hline
4.0&       -0.6955506394&    -4.025940635& 0.4300576700&    0.5304281261 \\
\hline
5.0&        -0.6772916132&    -4.941274459& 0.4476172910&    0.5276798434\\    
\hline
6.0&        -0.6573105590&    -5.903659889& 0.4571245682&    0.5253680730\\  
\hline
7.0&        -0.6391288554&    -6.890997919& 0.4617645750&    0.5234038890\\ 
\hline
8.0&       -0.6236060156&    -7.890707161&  0.4636608160&    0.5216941120 \\    
\hline
9.0&      -0.6106549406&    -8.895880333&   0.4641349950&    0.5201842000 \\    
\hline
10.0&     -0.5999010686&    -9.902954530&   0.4639483020&    0.5188413890 \\ 
\hline
15.0&      -0.5667087290&   -14.93326111&  0.4616835060&    0.5139612540 \\    
\hline
20.0&      -0.5500141977&   -19.94996191&  0.4603986900&    0.5109975560 \\    
\hline
25.0&      -0.5400058003&   -24.95998445&  0.4597076900&    0.5090476900 \\ 
\hline   
\end{tabular}
\end{table}